\documentstyle[12pt]{article}

\begin{document}

\title{NUCLEAR BROADENING OF\\ OUT--OF--PLANE TRANSVERSE MOMENTUM\\
IN DI-JET PRODUCTION\\}

\author{ Boris Z.~Kopeliovich
\thanks
{Address after August 15, 1994:  Universit\"at
Heidelberg,
Institut f\"ur \newline Theoretische Physik,
Philosophenweg 19, D-69120 Heidelberg, Germany.
\newline E-mail:  boris@bethe.npl.washington.edu}
\\
Joint Institute for Nuclear Research\\
Head Post Office, P.O. Box 79, 101000 Moscow,
Russia\\}
\date{}

\maketitle

\begin{abstract}

Recently claimed \cite{dj1,dj2,FC} anomalous nuclear effects in di-jet
production are analyzed in view of multiple interaction of
projectile/ejectile partons in nuclear matter.	We derive
model--independent relations between A-dependence of the cross section
and nuclear broadening of transverse momentum.	Comparison with the
data show that initial/final state interaction of partons
participating in hard process is hard as well.	This is a solid
argument in favor of smallness of a color neutralization radius of a
hadronizing highly virtual quark.

\end{abstract}
\vspace{2cm}

\begin{center}
Talk presented at the 5th Conference on the\\
Intersection of Particle and Nuclear Physics\\
St. Petersburg, Florida, 1994
\end{center}
\newpage

\section*{INTRODUCTION}

Nuclei are unique analysers of hadronization at early stage.  Initial
and final state interaction with nuclear environment modifies both
longitudinal \cite{Kop} and transverse \cite{Brod,DHK} momenta of
produced particles.  A well known example is the effect of nuclear
broadening of transverse momentum of Drell-Yan lepton pairs.
Experimental data \cite{DY} demonstrate an approximate proportionality
of the broadening, $\delta\langle k_T^2\rangle$, to the average path
of a quark in nuclear matter, what is a clear signal of importance of
multiple interactions.	This is supported by parameter--free
perturbative QCD calculations \cite{DHK}, which well agree with the
data.  Numerically nuclear effects are rather small, $\delta\langle
k_T^2\rangle\approx 0.15\;GeV^2$ for heavy nuclei, as compared with
$k_T^2\approx 1\;GeV^2$ on a proton target.

Much bigger effect was observed by the E609 Collaboration \cite{dj1}
in production of di-jets with $p_T>4\;GeV$ by $400\;GeV$ protons.  The
broadening of out-of-plane transverse momentum squared, $\delta\langle
k_{T}^2\rangle\approx 2-3\;GeV^2$ on heavy nuclei, is more than an
order of magnitude larger than that in the Drell-Yan reaction.	A
smaller effect, $\delta\langle k_{T}^2 \rangle\approx 1\;GeV^2$ was
observed in experiment E683 \cite{dj2} in photoproduction of di-jets
with transfer momenta, $p_T>3\;GeV$.  Nevertheless, in both cases
nuclear effects substantially exceed what is known for Drell-Yan
reaction.  Although the whole path of projectile and ejectile partons
through nuclear matter in di--jet production is three
times as long as in the Drell-Yan case, this factor alone
cannot explain the difference in $\delta\langle k_{T}^2 \rangle$.

Another manifestation of multiple interactions in nuclear matter, so
called Cronin effect, is known since 1973 \cite{Cronin}.  An exponent
$\alpha$, characterizing $A$-dependence of the cross section of
inclusive hadron production, parameterized as $A^{\alpha}$, increases
with the hadron transverse momentum, starting from
$\alpha\approx 2/3$ at small $p_T$ up to $\alpha >1$ at high $p_T$ (a
few $GeV/c$).  The growth of $\alpha$ mirrors the increase of the
interaction multiplicity, since each rescattering adds roughly
$1/3$ to $\alpha$ due to integration over longitudinal coordinate.

This sheds light on the mechanism of anomalous nuclear broadening of
transverse momentum in di-jet production.  The more the partons
interact with nuclear medium, the larger is the disbalance, $k_T^2$,
of the jet transverse momenta.	Therefore $\delta\langle
k_{T}^2\rangle$ can substantially exceed the value, which one could
naively expect on analogy with Drell-Yan reaction.

In this paper we try to establish in a least model--dependent way a
relation between the nuclear antishadowing at high $p_T$ ($\alpha>1$)
and the broadening of the out--of--plane transverse momentum, $\langle
k_{T}^2\rangle$, in di--jet production.  Using this relation and
available data on di--jet production on nuclei, we come to the
conclusion that interaction of a highly virtual parton cannot be
soft, what signifies a smallness of the color neutralization radius.

\section*{NUCLEAR BROADENING OF TRANSVERSE MOMENTA VERSUS A-DEPENDENCE
OF CROSS SECTION}

\subsection*{\large\bf Single jet production}
The cross section of a parton--nucleus scattering with high $p_T$ can
be written in the form,

\begin{equation}
\frac{d\sigma^A}{d^2p_T}=\int d^2b\sigma^A(p_T,b),
\label{1}
\end{equation}
where the partial cross section $\sigma^A(p_T,b)$ at impact parameter
$b$ reads,

\begin{equation}
\sigma^A(p_T,b)=\sum_{n=1}^{A}\frac{T^n}{n!}\;e^{-\sigma_0T}
\int\prod_{i=1}^nd^2k_i\;\frac{d\sigma}{d^2k_i}
\;\delta\left(\sum_{j=1}^n\vec k_j - \vec p_T\right)
\label{1a}
\end{equation}

Here $T\equiv T(b)=\int_{-\infty}^{\infty}\;dz\;\rho_A(b,z)$ is the
nuclear thickness.  The nuclear density is normalized as, $\int d^3r
\rho_A(\vec r)=A$.  The factor $T^n/n!$ originates from integration
over longitudinal coordinates, $z_i$, of bound nucleons, participating
in the multiple interaction of the parton.  $d\sigma/d^2k$ is
differential cross section of scattering of the parton on a nucleon
with transverse momentum transfer $\vec k$, summed over final states.
The factor $\exp(-\sigma_0\;T)$ takes into account the condition that
no more except $n$ rescatterings occurs; $\sigma_0=\int d^2k\;
(d\sigma/d^2k)$ is the total parton--nucleon interaction cross
section.

It is convenient to switch in (\ref{1a}) to impact
parameter representation.

\begin{equation}
\sigma^A(p_T,b)=e^{-\sigma(0)T}\int\frac{d^2\rho}{(2\pi)^2}\;
e^{i\vec p_T\vec\rho}
\left[e^{\sigma(\rho)T}-1\right] ,
\label{2}
\end{equation}
where

\begin{equation}
\sigma(\rho)=\int d^2k\;e^{i\vec k\vec\rho}\frac{d\sigma}{d^2k} \;,
\label{3}
\end{equation}
and $\sigma(0)\equiv \sigma_0$.

Expression (\ref{2}) is model--independent, all dynamics is hidden
in $\sigma(\rho)$.

Using (\ref{1a}) one calculates the average number of rescatterings
at impact parameter $b$,

\begin{equation}
\langle n(b)\rangle=\frac{1}{\sigma^A(p_T,b)}\;
e^{-\sigma(0)T}\int\frac{d^2\rho}{(2\pi)^2}\;e^{i\vec p_T\vec\rho}\;
\sigma(\rho)T\;e^{\sigma(\rho)T}
\label{4}
\end{equation}

According to eq. (\ref{1a}) each interaction of the parton during
propagation through the nucleus provides an extra factor $T$.
Thus $T$-dependence
of $\sigma^A(p_T,b)\propto T^{\beta}$ correlates with $\langle
n(b)\rangle$.  Using (\ref{2}) and (\ref{4}) one gets
$\beta=d\;ln[\sigma^A(p_T,b)]/d\;ln(T)$,

\begin{equation}
\beta=\langle n(b)\rangle - \sigma(0)\;T
\label{5}
\end{equation}

Since $T(b)\propto A^{1/3}$ and the integration over impact parameter,
$b$, provides an extra factor $A^{2/3}$, we arrive at the relation
between the mean number of multiple interactions in a nucleus and the
exponent $\alpha$, characterizing $A$ dependence of
$d\sigma^A/d^2p_T\propto A^{\alpha}$,

\begin{equation}
\langle n\rangle=\gamma\;\sigma(0)\langle T\rangle+3\alpha-2\;,
\label{6}
\end{equation}
where $\gamma=1$; $\langle T\rangle={1\over A}\int d^2b\;T^2(b)$.

The interpretation of relation (\ref{6}) is transparent.  At small
transverse momenta $\alpha\approx 2/3$, so $\langle
n\rangle=\sigma(0)T$, what correspond to the usual definition of mean
multiplicity of projectile interactions in Glauber model.  This
term contributes
at high $p_T$ as well, however on top of that $\langle n\rangle$ may
grow with $p_T$ if
additional rescatterings enlarge the cross section.
This depends on details of the dynamics, which we try to avoid,
connecting
$\langle n\rangle$ with experimentally measured $\alpha$. For
instance,
$\alpha=1$ means that $\langle n\rangle=1+\sigma_0T$, i.e. there is
one hard scattering and $\sigma_0T$ soft rescatterings along the
trajectory of the parton. In the case of nuclear antishadowing,
$\alpha>1$, the number of rescatterings increases in accordance with
(\ref{6}).

As a result of multiple interaction in nuclear matter
a parton gains an additional transverse momentum,

\begin{equation}
\delta\langle k_T^2\rangle=
2\gamma\;\delta\langle k_T^2\rangle_{DY} +
k_0^2(3\alpha-2)
\label{6a}
\end{equation}

Here $k_0^2$ is a mean square of momentum transfer in each
rescattering of the parton.  $\delta\langle k_T^2\rangle_{DY}={1\over
2}k_0^2\sigma(0)\langle T\rangle$ is the same as the nuclear
broadening of transverse momentum in Drell-Yan process.  It was argued
in \cite{DHK} that $\delta\langle k_T^2\rangle_{DY}$ is
$k_0^2$--independent.
Indeed, due to confinement a parton color is neutralized by
accompanying partons.  If $r_s$ is a radius of color neutralization in
the transverse plane, color screening cuts off soft interactions, so
$k_0^2\approx 1/r_s^2$.  On the other hand, $\sigma(0)\propto r_s^2$
due to color transparency.  Thus $\delta\langle k_T^2\rangle_{DY}$ is
$r^2_s$--independent \cite{DHK}.

At first sight relation (\ref{6a}) contradicts data. Indeed, $\alpha$
is usually $A$--independent or decreases with $A$. On the other hand,
$k_T^2$ rises with $A$ \cite{dj1,dj2}. However, the derivation of
(\ref{3}), (\ref{4}) silently assumed $n\ll A$, what is true only for
heavy nuclei.

Thus, one can use (\ref{6a}) for heavy nuclei to get information about
magnitude of $k_0^2$, characterizing hardness of initial/final state
interaction with nuclear medium.  Unfortunately in the case of
single--jet production the in--plane broadening of transverse momentum
is affected by the trigger--bias effect.  In order to study the
out--of--plane broadening one needs a recoil particle or a jet, in
order to fix the scattering plane.

 \subsection*{\large\bf Di-jets}

In the case of back--to--back di-jet production a
projectile parton produces in a hard interaction a
pair of partons with transverse momenta $\vec Q_1$ and $\vec Q_2$,
initiating the jets.  In the case of interaction with a nucleus
multiple interactions of projectile and ejectile partons affect the
cross section $d\sigma^A/d^2p_1d^2p_2$, increasing the disbalance of
transverse momenta. The corresponding partial cross section reads,

\begin{eqnarray}
\sigma^A(p_1,p_2,b)&=&\sum_{l,m,n=0}^{A-1}
\int_0^Tdt\;\frac{t^l(T-t)^{m+n}}{l!m!n!}\;
e^{-\sigma_0(2T-t)}
\int d^2Q_1d^2Q_2\;\frac{d\sigma^N}{d^2Q_1d^2Q_2}
\nonumber\\
& &\int\prod_{i=0}^ld^2k_i\;\frac{d\sigma}{d^2k_i}
\int\prod_{j=0}^md^2q_{1j}\;\frac{d\sigma}{d^2q_{1j}}\;
\delta\left(\sum_{i=0}^l\vec k_i +\vec Q_1+\sum_{j=0}^m\vec q_{1j} -
\vec p_1\right)
\nonumber\\
& & \int\prod_{k=0}^nd^2q_{2k}\;\frac{d\sigma}{d^2q_{2k}}\;
\delta\left(\sum_{i=0}^l\vec k_i+\vec Q_2+\sum_{k=0}^n\vec q_{2k}-
\vec p_2\right) 	   ,
\label{7}
\end{eqnarray}

where $t$ is a nuclear thickness covered by the projectile parton
before the hard collision.  $d\sigma^N/d^2Q_1d^2Q_2$ is a cross
section of back-to-back di--parton production in the projectile parton
-- nucleon interaction.  Note, that in the laboratory frame it cannot
be treated as a high-$p_T$ parton-parton elastic scattering, since it
is forbidden by kinematics.  It should be treated as a partonic
fluctuation, containing prepared parton pair with intrinsic
high-$p_T$, which is released as a result of interaction with a
target.

Expression (\ref{7}) has much simpler form in impact-parameter
representation,

\begin{eqnarray}
& &\sigma^A(p_1,p_2.b)=\int\frac{d^2\rho_1d^2\rho_2}{(2\pi)^4}\;
\frac{\sigma^N(\rho_1,\rho_2)\;
\exp\left[i(\vec p_1\vec\rho_1+\vec p_2\vec\rho_2)\right]}
{\sigma(\rho_1)+\sigma(\rho_2)-
\sigma(\rho_1+\rho_2)-\sigma(0)} \nonumber\\
\nonumber\\
& &\left[\exp\left\{\left[\sigma(\rho_1)+
\sigma(\rho_2)-2\sigma(0)\right] T \right\} -
\exp\left\{\left[\sigma(\rho_1+\rho_2)-
\sigma(0)\right]T\right\}\right]     ,
\label{8}
\end{eqnarray}
where

\begin{equation}
\sigma(\rho_1,\rho_2)=
\int d^2Q_1d^2Q_2\;e^{i\vec Q_1\vec\rho_1+i\vec Q_2\vec\rho}
\;\frac{d\sigma^N}{d^2Q_1d^2Q_2}
\label{9}
\end{equation}

Unfortunately no exact relation between total multiplicity of
interactions the exponent $\alpha$, characterizing A-dependence,
follows from (\ref{7})--(\ref{8}), but only an approximate one.  It
can be represented in the same form (\ref{6}), where $\langle
n\rangle$ is the total number of interactions of the projectile and
ejectile partons, excluding the hard interaction initiating the
observed di-jet. The uncertainty is contained in
The factor $\gamma$, is model-dependent in this case, but is
restricted by
$1<\gamma<2$. Nevertheless, this does not bring much uncertainty to
nuclear broadening of the out-of-plane transverse momentum,

\begin{equation}
\delta\langle k_T^2\rangle=
2\gamma\;\delta\langle k_T^2\rangle_{DY} +
3k_0^2(\alpha-1)
\label{10}     ,
\end{equation}
because the first term is relatively small.  Thus the accuracy of
(\ref{10}) is about $\pm 0.15\;GeV^2$ for heavy nuclei, whereas the
whole nuclear contribution, $\delta\langle k_T^2\rangle$, is a few
$GeV^2$ \cite{dj1,dj2}.

Using available experimental information for $\delta\langle
k_T^2\rangle\approx 2-3~GeV^2$ \cite{dj1} and $\alpha=1.18\pm 0.5$
\cite{Fields}, we estimate the mean momentum transfer squared, in a
single rescattering of a parton in nuclear matter, $k_0^2\geq
2~GeV^2$.  This value exceeds even the out-of-plane transverse
momentum squared, $\langle k_T^2\rangle\approx 1~GeV^2$, detected on a
proton target \cite{dj1}, what looks puzzling.	However the results of
analyses of di-jet events is extremely sensitive to used kinematical
cuts, background of underlying events, jet finding algorithm, etc.
\cite{dj1,dj2}.

Unfortunately the E683 Collaboration did not release yet their results
for A--dependence of the cross section.  Since this experiment has
lower bottom bound on $p_T$ of jets, than that used in \cite{dj1}, one
may expect that $\alpha$ is slightly smaller as well.  In such a case
the result of \cite{dj2} provide $k_0^2\geq 1~GeV^2$,

\section*{CONCLUSIONS. RADIUS OF COLOR NEUTRALIZATION}

We arrive at a conclusion that the Cronin effect ($\alpha>1$)
\cite{Cronin} and the anomalous nuclear broadening of the out-of-plane
transverse momentum \cite{dj1,dj2,FC} are closely related phenomena.
Namely the expression (\ref{10}) connects these two.  The closeness of
the measured value of $\alpha$ to $1$ points at rareness of
initial/final state interaction with nuclear matter of participating
partons.  On the other hand, the nuclear broadening of transverse
momenta is substantial.  This proves our claim that partons
participating in hard reaction interact hardly.  The mechanism, which
cut of soft interaction of highly virtual quarks, is color screening.
Due to confinement a colored parton is always accompanied with other
partons neutralizing its color.  If the radius of color neutralization
were small in transverse plane, color screening would cut off soft
gluon exchanges, making reinteractions hard.  If it is true or not
depends on poorly known confinement dynamics.	An appropriate model
in the case of highly virtual quarks is color neutralization by gluon
bremsstrahlung \cite{Gribov}.  The transverse momenta of gluons
emitted as a result of a high-$p_T$ scattering are of the order of
$p_T$.	Hence the radiation is located around the parton with
transverse separation $r\approx
1/p_T$.  This confirms the above idea about smallness of the color
neutralization radius of highly virtual partons, and nicely explains
both the rareness of parton reinteraction in nuclear matter and large
broadening of transverse momentum.  The hardness of the initial/final
state interaction in hard reactions is a phenomenon analogous to color
transparency.

This conclusion differs from the assumption \cite{LQS}, that
the highly virtual partons interact softly.  The authors calculated in
PQCD the single-rescattering correction to di-jet photoproduction,
provided a broadening of transverse momentum, proportional to
$A^{1/3}$.  However, they did not pretend for explanation of the data,
since calculated the correction only up to an unknown factor,
$\lambda^2$, having dimension of mass squared.	This parameter was
adjusted to fit the data \cite{dj1,dj2} at $\lambda^2\approx
0.05-0.1~GeV^2$.  As different from the author's guess that
$\lambda^2\approx \Lambda_{QCD}^2$, it follows from present
consideration that $\lambda^2\approx \rho_Ar_0$, where $\rho_A\approx
0.16~Fm^{-3}$ is the nuclear density; $r_0\approx 1.2~Fm$ is a factor
in the formula, describing $A$-dependence of nuclear radius,
$R_A\approx r_0A^{1/3}$.  This gives $\lambda^2\approx 0.008$, much
less than one needs, according to \cite{LQS}, to explain the data.

As soon as the radius of color neutralization, $r_s$ is small, the
initial/final state interactions should be insensitive to a deconfined
environment (quark-gluon plasma), if $r<r_D$, where $r_D$ is the Debye
screening radius. Debye mass squared, predicted at energies of LHC is
$\mu_D^2\approx 0.4~GeV^2$, what is less than our estimate of
the parameter $k_0^2\geq 1~GeV^2$ from data \cite{dj1,dj2}. Besides,
$k_0^2$ is expected to grow with $p_T$.  This observation makes
questionable a effectiveness of hard probes of quark-gluon plasma, for
instance jet quenching due to Debye screening \cite{Gyulassy}.	Energy
loss for induced gluon radiation, according to our results, is to be
proportional to $k_0^2$, rather than to $\mu^2_D$, i.e.  is the same
in cold nuclear matter and quark-gluon plasma.\medskip

I appreciate useful discussions with Stan Brodsky, who drew my
attention to the problem under consideration, Marj Corcoran and
Xin-Nian Wang.

\end{document}